\documentstyle[art12]{article}
\begin{document}
\begin{center}
{\large\bf $B_c$ meson production in $\gamma\gamma$ collisions\\[2mm]
and charm content of the photon}\\
\vspace{4mm}
{\it A.P.Martynenko and V.A. Saleev}\\
\vspace{4mm}
 Samara State University, Samara, 443011, Russia
\end{center}

\begin{abstract}

Based on photon structure function formalism we have calculated
$B_c(B_c^{*})$ meson production cross section in $\gamma\gamma$
collisions via partonic subprocess $c\gamma\to B_c(B_c^{*}) b$. It was
shown that this approach gives the same results for $B_c^*(B_c)$ meson
cross sections as the direct leading order QCD calculation of the
subprocess $\gamma\gamma\to B_c(B_c^{*})b\bar c$ at small energy.
Howere, opposite to the direct QCD calculation we have obtained the
increase of the $B_c(B_c^{*})$ meson cross section at large energies.
We have found also additional contribution to the $B_c$ meson
production via the nonperturbative fluctuations $\gamma\leftrightarrow
J/\psi$.

\end{abstract}

The recent  measurements  at  HERA  $ep$   collider   \cite{1}   shown
nontrivial role of the photon structure function (PSF) \cite{1b} in
photoproduction
of jets and charm quarks at high energies. Usually it is supposed that
the  gluon  content of the photon in the resolved photon interaction
is dominant
at high energies \cite{2}.  However,  the charm content of the
photon   may   be   also   very   important  for  processes  of  charm
photoproduction as well as the charm content of  the  proton  for  charm
hadroproduction    processes.    First   of   all,   the hadron-like
(nonperturbative) charm quark contribution is needed  for  description
of  the  large diffractive charm production \cite{3}.  At second,  the
many aspects of charm production in  the  hard  processes,  where  QCD
scale  parameter  $Q^2$ is large ($Q^2>>m_c^2$),  can be understood in
the context of perturbative QCD as a initial  state  charm  excitation
processes \cite{4}. Recently, we have calculated the cross sections of
the $J/\psi$ photo- and hadroproduction at the large transverse
momentum via charm quark excitation in
proton  \cite{5} and photon \cite{6}.  In the present paper we discuss
the contribution of the charm quark PSF in the $B_c(B_c^{*})$ meson
production via partonic subprocess $\gamma c\to B_c(B_c^{*})$
in $\gamma\gamma$ collisions. Because of the QCD scale parameter $Q^2$
in  this  process  is order to square of $B_c$ meson mass ($Q^2\approx
M_{B_c}^2>>m_c^2$),  the application of the $c$-quark PSF approach  for
$B_c(B_c^{*})$ meson production is well grounded as for
calculation of the large $p_{\bot}$ $B_c$ meson production as for the
calculation of the total cross section.
Taking into account the mass of the initial $c$-quark in the kinematic
relations as well as in the $c$-quark distribution functions, we can
control the mass corrections near the threshold of the $B_c$ meson
production.

The production    mechanisms    for   $B_c(B_c^{*})$   mesons   in
$\gamma\gamma$ collisions  have  been  discussed   early   in   Refs.
\cite{7,8}.   It   was   assumed   that  in  $\gamma\gamma$ collision
$B_c(B_c^{*})$ mesons are dominantly produced in association  with
a $b$- and $c$- quark jets via the partonic subprocess:

\begin{equation}
 \gamma+\gamma\to B_c(B_c^{*})+b+\bar c,
\end{equation}
which is described by twenty Feynman diagrams.
In this paper, we examine a $B_c$ and $B_c^{*}$ meson production
via the partonic subprocess:

\begin{equation}
c(q)+\gamma(k)\to B_c(P)+b(q_1).
\end{equation}
The above subprocess is described only by four Feynman diagrams
which are shown in Fig.1. The corresponding amplitudes can be expressed
as follows:
\begin{equation}
{{\cal M}_1}=e_beg^2T^aT^a\bar U_b(q_1)\hat\varepsilon_{\gamma}(k)
\frac{\hat q_1-\hat k+m_b}{(q_1-k)^2-m_b^2}\gamma_{\rho}
\frac{\hat\Pi}{(p_c-q)^2}\gamma^{\rho}U_c(q)
\end{equation}
\begin{equation}
{{\cal M}_2}=e_ceg^2T^aT^a\bar U_b(q_1)\gamma_{\rho}
\frac{\hat\Pi}{(p_b+q_1)^2}\gamma^{\rho}
\frac{\hat k+\hat q+m_c}{(q+k)^2-m_c^2}
\hat\varepsilon_{\gamma}(k)U_c(q)
\end{equation}
\begin{equation}
{{\cal M}_3}=e_ceg^2T^aT^a\bar U_b(q_1)\gamma_{\rho}
\frac{\hat\Pi}{(p_b+q_1)^2}\hat\varepsilon_{\gamma}(k)
\frac{\hat p_c-\hat k+m_c}{(p_c-k)^2-m_c^2}
\gamma^{\rho}U_c(q)
\end{equation}
\begin{equation}
{{\cal M}_4}=e_beg^2T^aT^a\bar U_b(q_1)\gamma_{\rho}
\frac{\hat k-\hat p_b+m_b}{(k-p_b)^2-m_b^2}\hat\varepsilon_{\gamma}(k)
\frac{\hat\Pi}{(p_c-q)^2}\gamma^{\rho}U_c(q)
\end{equation}
Here, $\bar  U_b(q_1)$  and  $U_c(q)$  denote the Dirac spinors of the
quarks,           $T^a=\lambda^a/2$,            $e=\sqrt{4\pi\alpha}$,
$g=\sqrt{4\pi\alpha_s}$, $e_c$ and $e_b$
 are electric charges of the quarks in units of $e$,  $m_c$ and  $m_b$
 are   quark   masses,   $\varepsilon_{\gamma}(k)$   is   the   photon
 polarization four-vector. The spin and colour properties of the $\bar
 b  c$  ground state are described in nonrelativistic approximation by
 following projection operator \cite{9}:

\begin{equation}
 \hat\Pi=\frac{F_c\Psi(0)}{2\sqrt{M}}\hat a(\hat P+M),
\end{equation}
where $F_c=\delta^{kr}/\sqrt{3}$,  $k$  and  $r$ are colour indexes of
$\bar b-$ and $c-$ quarks,  $M=m_c+m_b$  is  $B_c$  meson
mass, $p_c=\frac{m_c}{M}P$ and $p_b=\frac{m_b}{M}P$ are $c-$ and $\bar
b-$  quarks  four-momenta,  $P=p_c+p_b$  is  $B_c$  meson
four-momentum,  $\hat  a=\gamma_5$  for  pseudoscalar $\bar bc$ ground
state and $\hat a=\hat\varepsilon$ for vector $\bar b c$ ground state.
The  $\Psi(0)$  is a nonrelativistic wave function at the origin which
can be presented by the corresponding decay constants:

 \begin{equation}
 f_{B_c}=f_{B_c^{*}}=\sqrt\frac{12}{M}\Psi(0).
\end{equation}
 We put in our calculation: $f_{B_c}=f_{B_c^{*}}=0.57$ GeV,
 $\alpha_s=0.2$, $\alpha=1/137$, $m_c=1.5$ GeV and $m_b=4.8$ GeV.

After average and sum over spins and colours of initial and final
particles, we have obtained for square of matrix element:
\begin{equation}
\overline{|{\cal M}|^2}=B_{\gamma c}\sum^4_{i,j=1}e_ie_j
F_{ij}(\hat s,\hat t,\hat u),
\end{equation}
 where
$$B_{\gamma c}=\frac{16}{27}\pi^3\alpha\alpha_s^2f_{B_c}^2,$$
$e_1=e_4=e_b$, $e_2=e_3=e_c$, $\hat s=(k+q)^2$, $\hat t=(k-q_1)^2$,
 $\hat u=(q-q_1)^2$ and $\hat s+\hat t+\hat u=m_c^2+m_b^2+M^2$.
The explicit analytical formulae for functions $F_{ij}$ are
obtained using the symbolic manipulation program FORM \cite{10}
and automatically implemented into a FORTRAN program.

The differential cross section for subprocess $c\gamma\to
B_c(B_c^{*})b$ can be written as follows:
\begin{equation}
\frac{d\hat \sigma}{d\hat t}=\frac{\overline{|{\cal M}|^2}}
 {16\pi(\hat s-m_c^2)^2}.
\end{equation}

For description of the $B_c(B_c^{*})$ meson photoproduction
via charm quark PSF we can use a  conventional  parton  model.
In this approach the measurable cross section is obtained by
folding the  hard
parton level cross section with the respective parton densities:

\begin{equation}
\sigma(\gamma\gamma\to B_c X)=\int_{x_{min}}^{x_{max}}dx C_{\gamma}(x,Q^2)
 \int_{\hat t_{min}}^{\hat t_{max}}\frac{d\hat\sigma}{d\hat t}
  (c\gamma\to B_c b),
\end{equation}
where $C_{\gamma}(x,Q^2)$ is a charm quark distribution of a photon,

$$\hat t_{max\atop{min}}=
    m_b^2-\frac{\hat s-m_c^2}{2\hat s}[\hat s+m_b^2-M^2\pm
    \sqrt{(\hat s-(m_b+M)^2)(\hat s-m_c^2)}],$$

\begin{equation}
\hat s=xs_{\gamma\gamma}+m_c^2,
\end{equation}
$$x_{min}=4m_bM/s_{\gamma\gamma},\qquad x_{max}=Q^2/(Q^2+4m_c^2)$$
and $s_{\gamma\gamma}$ is the
square of a total energy of photons in the $\gamma\gamma$
center-of-mass reference frame.

The charm content of a photon can be presented as the composition of a
hadron--like PSF and a point--like (quark-gluon) PSF \cite{1b}:
\begin{equation}
C_{\gamma}(x,Q^2)=C_{\gamma}^{had}(x,Q^2)+C_{\gamma}^{pl}(x,Q^2).
\end{equation}
In accordance with the vector-meson-dominance (VDM) model \cite{12},
the hadron--like $c$-quark PSF is presented via the charm quark
distribution function of $J/\psi$ meson:
\begin{equation}
C_{\gamma}^{had}(x,Q^2)= k\frac{4\pi\alpha}{f_{\psi}^2}C_{\psi}(x,Q^2)
\end{equation}
 with $1\leq k\leq 2$. The precise value of $ k$ clearly has
 to be extracted from experiment.  Similar to Ref.\cite{13}, where
  the photoproduction of charm hadrons
  in VMD model have been discussed,
 we used for $C_{\psi}(x,Q^2)$ the simple scaling parametrization,
which takes into account c-quark mass effects:
 \begin{equation}
  C_{\psi}(x)=49.5x^{2.2}(1-x)^{2.45}.
\end{equation}

The $c$-quark point-like part of the PSF can be calculated using
perturbative approach. As it was shown in Ref.\cite{12}, the distribution
of a heavy quark with mass $m_q$ in the photon near the threshold
($\sqrt{s_{\gamma\gamma}}\geq 2m_q$) should be calculated from
the complete massive leading order (LO) Bethe - Heitler cross section for
$\gamma^*\gamma\to q\bar q$ reaction:

\begin{equation}
C_{\gamma}^{pl}(x,Q^2)=\frac{3\alpha}{2\pi}e_c^2F(x,\frac{m_c^2}{Q^2}),
\end{equation}
where
\begin{eqnarray}
&&\qquad F(x,r)=\beta[-1+8x(1-x)-4rx(1-x)]+\nonumber\\
&&[x^2+(1-x)^2+4rx(1-3x)-8r^2x^2]\ln[(1+\beta)/(1-\beta)],\nonumber
\end{eqnarray}
$$\beta=\sqrt{1-4rx/(1-x)}.$$


Recently, the next leading order (NLO) corrections for the $c$-quark
PSF was computed in Ref. \cite{14}. Effectively, the point-like part
of the $c-$quark PSF may be presented as a sum of Bethe -- Heitler
formula (16) (parametrization BH) and, so-called hadronic component
\cite{14}, which corresponds to $c-$quark production via QCD
evolution of the PSF.

In the Fig.2 $c$-quark PSF's are presented at scale $Q^2=M^2$. The
shapes of the curves for VMD and massive LO point-like parametrization
(BH) are the same. Opposite these ones, point-like hadronic component
of the $c-$quark PSF (parametrization PLH) increase at $x\to 0$. Below
$x$ about $0.01$ the $c-$quark PSF is completely dominated by the
point-like hadronic component. Obviously, that the following
$x-$dependence gives the growth of the $B_c$ meson $\gamma\gamma-$
production cross section at large $\sqrt{s_{\gamma\gamma}}>>M$.


The results of our calculation for $B_c^{*}$ meson production
cross   section   in   $\gamma\gamma$   collisions   as a function  of
$\sqrt{s_{\gamma\gamma}}$ are shown in Fig. 3.  Curves 1 - 3
are  the  PSF  contributions  via  subprocess $c\gamma\to
B_c^{*}b$  with  BH, VMD  and  PLH  parametrization of PSF,
respectively.  Stars show  the
contribution of the direct production subprocess $\gamma\gamma\to
B_c^{*} b\bar c$ and was  taken from Ref.\cite{7}.  Note,
that curve 2 was obtained at $k=1$.
One can see that BH parametrization of the $c$-quark PSF, based on the
Bethe-Heitler formula, well coincides to the direct leading order QCD
calculation from Ref.\cite{7} as near the peak of the $B_c^{*}$
meson production as near threshold, i.e. at $\sqrt{s_{\gamma\gamma}}=
 15\div 40$ GeV. The contribution of the hadron-like part of the
$c$-quark PSF is about 30\% (at $k=1$) of the point-like part
contribution at the $\sqrt{s_{\gamma\gamma}}=15\div 25$ GeV. At the
large $\sqrt{s_{\gamma\gamma}}$ VMD contribution speedily decrease.
Howere, if we take in mind, that for $c$-quarks as well as for light
ones, it has from the experiment $k\approx 2$, we have obtained
large additional contribution to the $B_c^{*}$ - meson production
via the nonperturbative fluctuations
$\gamma\leftrightarrow J/\psi$.

The contribution of the point-like hadronic component is very small at
$\sqrt {s_{\gamma\gamma}}\le 50$ GeV.
But at large energies the curve 3 demonstrates us nonfalling versus
energy $B_c^{*}$ meson production cross section. So, at
$\sqrt{s_{\gamma\gamma}}=100$ GeV the total resolved photon contribution
five times the contribution of the direct
production mechanism.

In Refs.\cite{7,8} the strong suppression of the pseudoscalar  state
  production  in comparison with the production of the vector one near
  threshold  was  obtained,  so at  $\sqrt{s_{\gamma\gamma}}=15$  GeV
  $\sigma_{B_c^{*}}/\sigma_{B_c}\sim 55$\cite{7} and at
  $\sqrt{s_{\gamma\gamma}}=20$ GeV
  $\sigma_{B_c^{*}}/\sigma_{B_c}\sim 15$\cite{7} or
  $\sigma_{B_c^{*}}/\sigma_{B_c}\sim 10$\cite{8}.  In spite of the
  fact  that  the  results  of our calculation for $B_c^{*}$ meson
  production (Fig.3) at  $\sqrt{s_{\gamma\gamma}}=15\div 40$  GeV  are  in
  agreement  with Refs.\cite{7,8},  the results for pseudoscalar $B_c$
  meson  production   are   very   different   in   this   region   of
  $\sqrt{s_{\gamma\gamma}}$.  We have obtained that the
  $\sigma_{B_c^{*}}/\sigma_{B_c}$ ratio is approximately equal  to
  3 at all energies.  We don't see the
  reason  of  strong  additional  suppression  of  the   $B_c$   meson
  production   in   comparison   with  $B_c^{*}$  production  near
  threshold in the partonic subprocesses $c\gamma\to B_c(B_c^{*})b$.

The differential distributions of  the  $B_c$  mesons  in
$\gamma\gamma$ collisions are also very interested. Figs. 4 and 5 show
the spectra of $B_c^{*}$ mesons versus the scaled energy  variable
$z=2E_{B_c^{*}}/\sqrt{s_{\gamma\gamma}}$    obtained    from   the
diagrams  of the subprocess (2) at
$\sqrt{s_{\gamma\gamma}}=20$   GeV  and  $\sqrt{s_{\gamma\gamma}}=100$
GeV, respectively. At small energy (Fig.4) the shapes of the
$z-$spectra obtained using different parametrization are the same and
the $B_c^{*}$ meson production in the central region of $z$ is
dominant.
The contribution of the point-like hadronic component at this energy is
negligibly small.
At high energy (Fig.5) the predictions of the discussed here $c-$quark
PSF components
are very different as for shape of the
$z-$spectra as for the absolute values. The VMD and BH parametrizations
predict two diffractive-like peaks both at $z\approx 0$ and $z\approx 1$
regions.
The curve 3 (Fig.5) shows $z$-spectrum which was obtained using PLH
parametrization.
We see that the contributions of the pure photonic (BH) and hadronic
(PLH) components of the PSF are dominant at different $z$ and they may
be separate experimentally.

Note, that the shape of the $z$-spectra in Figs.4 and 5 are remarkable
different from result which was obtained in
Ref.\cite{8}  using  the  model  of the direct $B_c$ meson production.
This fact gives us additional opportunity to separate the contribution
of the resolved-photon interaction process in $B_c^{*}$ meson
$\gamma\gamma$-production already at not very large energies
$\sqrt{s_{\gamma\gamma}}=20-50$ GeV.

Really, the  generation  of  fixed energy $\gamma\gamma$-beams is very
difficult problem.  Now  we  may  study  $\gamma\gamma$-collisions  at
$e^+e^-$machines using the Weizsacker--Williams bremsstrahlung photon
spectrum \cite{16} or using the photon  spectrum  obtained  by  Compton
back-scattering  of  the  laser photons on high energy $e^+e^-$ beams
\cite{17}. Fig. 6 shows the results of our calculation for
$B_c^{*}$  meson  production  cross section in $e^+e^-$
collisions versus the $e^+e^-$ center-of-mass energy $\sqrt{s}$.  The
curves  in  Fig.5 are obtained by folding $\gamma\gamma$ cross section
with Weizsacker--Williams bremsstrahlung photon spectrum (lower curves)
and with  the  spectrum of the back-scattered laser
photons (upper curves).
The cross section of $B_c^{*}$ meson production from
bremsstrahlung  photons  is  very  small at LEP energies but increases
logarithmically with energy as the same Ref.\cite{8}.
We see that the
contribution  of the point-like hadronic component of the $c-$quark PSF
becomes important only in the TeV energy
range.  In contrast,  the cross section  of  $B_c^{*}$  meson
production  from  Compton  back-scattered photons has a maximum at LEP
energies.  The  energy  dependence  above
$\sqrt{s}=100$  GeV
predicted  by  the  model of the direct production \cite{8} and by the
resolved photon mechanism are very  different.  The  cross  sections
obtained in Ref.\cite{8}, as well as in this paper for VMD and BH
approaches,
fall  at  $\sqrt{s}>100$  GeV,  but   the   $B_c^{*}$   meson
production cross section, which was calculated using PLH
parametrization and taken into account NLO QCD corrections,
speedily
increases at  $\sqrt{s}>100$  GeV.  So,  at $\sqrt{s}=1$ TeV one has
$\sigma_{B_c^{*}}\sim
0.4$ pb.  The  values of the cross sections in the peak of
$B_c(B_c^{*})$ meson  $e^+e^-$-production  are
$\sigma_{B_c^{*}}\approx 0.2$ pb,  $\sigma_{B_c}\approx 0.06$ pb.
At  the  TeV  energy  region  one  has  large  contribution   of   the
resolved photon   interaction:  $\sigma_{B_c^{*}}\approx  1$  pb,
$\sigma_{B_c}\approx 0.3$ pb.

Based on photon structure function formalism we have calculated
$B_c(B_c^{*})$
meson production cross section in $\gamma\gamma$-collisions via
partonic subprocess $c(\gamma)\gamma\to B_c(B_c^{*}) b$.
It was shown that at the small energies this approach gives
the same results for
$B_c^{*}$ meson cross sections as the direct leading order QCD
calculation of the subprocess $\gamma\gamma\to B_c^{*}b\bar c$
at small energy.
Howere, opposite to the direct QCD calculation we have obtained the
increase of the $B_c(B_c^{*})$ meson cross section at large
energies. We
have found also additional contribution to the $B_c$ meson production
connected with the nonperturbative fluctuations
$\gamma\leftrightarrow J/\psi$.
It  was  shown  that  charm content of the photon
 may  be  study  experimentally  in  the
$B_c(B_c^{*})$  meson  production
already at the total energy 50 GeV in $\gamma\gamma$-collisions and at
the  total  energy  200  GeV  in  $e^+e^-$collisions  using  Compton
back-scattered photons.  The discussed here  $B_c(B_c^{*})$  meson
production   mechanism   can   be   used   for   prediction   of   the
$B_c(B_c^{*})$   meson   production   rates  at  high energies
$ep$ and $p\bar p$-colliders via partonic subprocess
$cg\to B_c(B_c^{*})b$ \cite{18}.

Authors thank A.K.~Likhoded for useful discussions of the $B_c$-meson
physics and E.\,Laenen for the valuable information about
obtained results.
This
research was supported by the Russian Foundation of Basic Research
(Grant 93-02-3545) and by State Committee on High Education of
Russian Federation (Grant 94-6.7-2015).

{\large\bf Figure captions.}

\begin{enumerate}
\item Diagrams used to describe the partonic subprocess
$c\gamma\to B_c(B_c^{*})b$.
\item C-quark distributions in the photon. Curves 1-3 are respectively
BH, VMD and PLH parametrizations.
\item Cross sections for $\gamma\gamma\to B_c^{*} X$ process
versus the c.m.
 energy $\sqrt{s_{\gamma\gamma}}$. Curves 1 - 3 are respectively
 contributions of $c\gamma\to B_c^{*}b$ subprocess using BH, VMD
 and PLH parametrizations.
The stars are the result of calculation $\gamma\gamma\to B_c^{*}
b\bar c$ subprocess, which is taken from Ref.\cite{7}.
\item Energy distribution of $B_c^{*}$ mesons at
$\sqrt{s_{\gamma\gamma}}=20$ GeV in the
 resolved photon interaction. The curves 1 and 2 are the same as Fig.3

\item Energy distribution of $B_c^{*}$ mesons at
$\sqrt{s_{\gamma\gamma}}=100$ GeV in the
 resolved-photon interaction. The curves are the same as Fig.3.
\item Cross sections for $e^+e^-\to B_c^{*} X$ process versus
the $e^+e^-$ c.m.
 energy $\sqrt{s}$ with bremsstrahlung (lower curves) and
 with back-scattered laser (upper curves) photon spectra.
 The curves are the same as Fig.3.
\end{enumerate}

\newpage
\def\varepsilonmline#1#2#3#4#5#6{%
       \put(#1,#2){\special{em:moveto}}%
       \put(#4,#5){\special{em:lineto}}}

\unitlength=1.00mm
\special{em:linewidth 1pt}
\linethickness{1pt}
\begin{picture}(130.00,94.20)
\varepsilonmline{24.00}{54.20}{1}{34.00}{64.09}{2}
\varepsilonmline{34.00}{64.09}{3}{34.00}{64.09}{4}
\varepsilonmline{34.00}{64.09}{5}{49.00}{64.09}{6}
\put(50.00,69.04){\oval(2.00,9.89)[]}
\varepsilonmline{50.00}{73.99}{7}{34.00}{73.99}{8}
\varepsilonmline{34.00}{73.99}{9}{34.00}{83.88}{10}
\varepsilonmline{34.00}{83.88}{11}{34.00}{83.88}{12}
\varepsilonmline{34.00}{83.88}{13}{51.00}{83.88}{14}
\varepsilonmline{34.00}{73.99}{15}{32.00}{71.84}{16}
\varepsilonmline{32.00}{71.84}{17}{34.00}{71.84}{18}
\varepsilonmline{34.00}{71.84}{19}{34.00}{71.84}{20}
\varepsilonmline{34.00}{71.84}{21}{32.00}{70.12}{22}
\varepsilonmline{32.00}{70.12}{23}{32.00}{70.12}{24}
\varepsilonmline{32.00}{70.12}{25}{34.00}{70.12}{26}
\varepsilonmline{34.00}{70.12}{27}{34.00}{70.12}{28}
\varepsilonmline{34.00}{70.12}{29}{32.00}{67.96}{30}
\varepsilonmline{32.00}{67.96}{31}{32.00}{67.96}{32}
\varepsilonmline{32.00}{67.96}{33}{34.00}{67.96}{34}
\varepsilonmline{34.00}{67.96}{35}{34.00}{67.96}{36}
\varepsilonmline{34.00}{67.96}{37}{32.00}{66.24}{38}
\varepsilonmline{32.00}{66.24}{39}{32.00}{66.24}{40}
\varepsilonmline{32.00}{66.24}{41}{34.00}{66.24}{42}
\varepsilonmline{34.00}{66.24}{43}{34.00}{66.24}{44}
\varepsilonmline{34.00}{66.24}{45}{32.00}{64.09}{46}
\varepsilonmline{32.00}{64.09}{47}{32.00}{64.09}{48}
\varepsilonmline{32.00}{64.09}{49}{34.00}{64.09}{50}
\varepsilonmline{51.00}{70.12}{51}{57.00}{70.12}{52}
\varepsilonmline{57.00}{70.12}{53}{51.00}{70.12}{54}
\varepsilonmline{51.00}{70.12}{55}{57.00}{70.12}{56}
\varepsilonmline{51.00}{67.96}{57}{57.00}{67.96}{58}
\varepsilonmline{56.00}{66.24}{59}{60.00}{68.82}{60}
\varepsilonmline{60.00}{68.82}{61}{60.00}{68.82}{62}
\varepsilonmline{60.00}{68.82}{63}{55.00}{71.84}{64}
\varepsilonmline{24.00}{94.20}{65}{26.00}{92.05}{66}
\varepsilonmline{28.00}{89.90}{67}{30.00}{88.18}{68}
\varepsilonmline{32.00}{86.03}{69}{34.00}{83.88}{70}
\put(32.00,91.19){\makebox(0,0)[cc]{k}}
\put(32.00,54.20){\makebox(0,0)[cc]{q}}
\varepsilonmline{92.00}{73.98}{71}{102.00}{73.98}{72}
\varepsilonmline{92.00}{73.98}{73}{82.00}{64.09}{74}
\varepsilonmline{82.00}{83.87}{75}{84.00}{82.15}{76}
\varepsilonmline{84.00}{82.15}{77}{84.00}{82.15}{78}
\varepsilonmline{86.00}{80.86}{79}{88.00}{79.14}{80}
\varepsilonmline{88.00}{79.14}{81}{88.00}{79.14}{82}
\varepsilonmline{90.00}{76.99}{83}{92.00}{74.84}{84}
\varepsilonmline{102.00}{73.98}{85}{104.00}{71.83}{86}
\varepsilonmline{104.00}{71.83}{87}{104.00}{73.98}{88}
\varepsilonmline{104.00}{73.98}{89}{106.00}{71.83}{90}
\varepsilonmline{106.00}{71.83}{91}{106.00}{73.98}{92}
\varepsilonmline{106.00}{73.98}{93}{108.00}{71.83}{94}
\varepsilonmline{108.00}{71.83}{95}{108.00}{73.98}{96}
\varepsilonmline{108.00}{73.98}{97}{110.00}{71.83}{98}
\varepsilonmline{110.00}{71.83}{99}{110.00}{73.98}{100}
\varepsilonmline{110.00}{73.98}{101}{112.00}{71.83}{102}
\varepsilonmline{112.00}{71.83}{103}{112.00}{73.98}{104}
\varepsilonmline{102.00}{73.98}{105}{109.00}{82.15}{106}
\varepsilonmline{112.00}{73.98}{107}{119.00}{82.15}{108}
\put(114.00,81.94){\oval(10.00,2.15)[]}
\varepsilonmline{112.00}{73.98}{109}{121.00}{67.10}{110}
\varepsilonmline{113.00}{83.01}{111}{118.00}{89.04}{112}
\varepsilonmline{116.00}{83.01}{113}{120.00}{88.18}{114}
\varepsilonmline{115.00}{88.18}{115}{121.00}{91.19}{116}
\varepsilonmline{121.00}{86.88}{117}{121.00}{91.19}{118}
\varepsilonmline{22.00}{17.63}{119}{50.00}{17.63}{120}
\varepsilonmline{33.00}{17.63}{121}{31.00}{19.78}{122}
\varepsilonmline{31.00}{19.78}{123}{33.00}{19.78}{124}
\varepsilonmline{33.00}{19.78}{125}{31.00}{21.93}{126}
\varepsilonmline{31.00}{21.93}{127}{33.00}{21.93}{128}
\varepsilonmline{33.00}{21.93}{129}{31.00}{23.65}{130}
\varepsilonmline{31.00}{23.65}{131}{33.00}{23.65}{132}
\varepsilonmline{33.00}{23.65}{133}{31.00}{25.80}{134}
\varepsilonmline{31.00}{25.80}{135}{33.00}{25.80}{136}
\varepsilonmline{33.00}{25.80}{137}{31.00}{27.52}{138}
\varepsilonmline{31.00}{27.52}{139}{50.00}{27.52}{140}
\put(50.00,22.58){\oval(2.00,9.89)[]}
\varepsilonmline{51.00}{23.65}{141}{57.00}{23.65}{142}
\varepsilonmline{51.00}{20.64}{143}{57.00}{20.64}{144}
\varepsilonmline{55.00}{25.80}{145}{60.00}{21.93}{146}
\varepsilonmline{55.00}{18.92}{147}{60.00}{22.79}{148}
\varepsilonmline{41.00}{27.52}{149}{39.00}{29.68}{150}
\varepsilonmline{38.00}{30.54}{151}{36.00}{32.69}{152}
\varepsilonmline{35.00}{33.55}{153}{33.00}{35.70}{154}
\varepsilonmline{32.00}{36.56}{155}{30.00}{38.71}{156}
\varepsilonmline{31.00}{27.52}{157}{50.00}{38.71}{158}
\varepsilonmline{93.00}{22.80}{159}{91.00}{24.95}{160}
\varepsilonmline{91.00}{24.95}{161}{91.00}{24.95}{162}
\varepsilonmline{91.00}{24.95}{163}{93.00}{24.95}{164}
\varepsilonmline{93.00}{24.95}{165}{93.00}{24.95}{166}
\varepsilonmline{93.00}{24.95}{167}{91.00}{27.10}{168}
\varepsilonmline{91.00}{27.10}{169}{91.00}{27.10}{170}
\varepsilonmline{91.00}{27.10}{171}{93.00}{27.10}{172}
\varepsilonmline{93.00}{27.10}{173}{93.00}{27.10}{174}
\varepsilonmline{93.00}{27.10}{175}{91.00}{28.82}{176}
\varepsilonmline{91.00}{28.82}{177}{93.00}{28.82}{178}
\varepsilonmline{93.00}{28.82}{179}{91.00}{30.97}{180}
\varepsilonmline{91.00}{30.97}{181}{93.00}{30.97}{182}
\varepsilonmline{93.00}{30.97}{183}{91.00}{32.69}{184}
\varepsilonmline{91.00}{32.69}{185}{91.00}{32.69}{186}
\put(112.00,27.75){\oval(2.00,9.89)[]}
\varepsilonmline{113.00}{28.82}{187}{120.00}{28.82}{188}
\varepsilonmline{113.00}{25.81}{189}{120.00}{25.81}{190}
\varepsilonmline{120.00}{25.81}{191}{120.00}{25.81}{192}
\varepsilonmline{118.00}{24.09}{193}{125.00}{27.10}{194}
\varepsilonmline{118.00}{30.97}{195}{125.00}{27.10}{196}
\varepsilonmline{93.00}{22.80}{197}{112.00}{22.80}{198}
\varepsilonmline{112.00}{32.69}{199}{82.00}{32.69}{200}
\varepsilonmline{102.00}{32.69}{201}{100.00}{34.84}{202}
\varepsilonmline{99.00}{35.70}{203}{97.00}{37.85}{204}
\varepsilonmline{96.00}{38.71}{205}{94.00}{40.86}{206}
\varepsilonmline{93.00}{41.72}{207}{91.00}{43.87}{208}
\varepsilonmline{93.00}{22.80}{209}{106.00}{13.77}{210}
\put(56.00,84.00){\makebox(0,0)[cc]{$q_1$}}
\put(67.00,69.00){\makebox(0,0)[cc]{$B_{c}(P)$}}
\put(44.00,78.00){\makebox(0,0)[cc]{$p_b$}}
\put(44.00,58.00){\makebox(0,0)[cc]{$p_c$}}
\put(125.00,64.00){\makebox(0,0)[cc]{$q_1$}}
\put(109.00,11.00){\makebox(0,0)[cc]{$q_1$}}
\put(54.00,39.00){\makebox(0,0)[cc]{$q_1$}}
\put(66.00,22.00){\makebox(0,0)[cc]{$B_{c}(P)$}}
\put(130.00,27.00){\makebox(0,0)[cc]{$B_{c}(P)$}}
\put(127.00,92.00){\makebox(0,0)[cc]{$B_{c}(P)$}}
\put(80.00,88.19){\makebox(0,0)[cc]{k}}
\put(87.00,43.19){\makebox(0,0)[cc]{k}}
\put(26.00,39.19){\makebox(0,0)[cc]{k}}
\put(24.00,12.20){\makebox(0,0)[cc]{q}}
\put(78.00,32.20){\makebox(0,0)[cc]{q}}
\put(79.00,62.20){\makebox(0,0)[cc]{q}}
\put(120.00,76.00){\makebox(0,0)[cc]{$p_b$}}
\put(106.00,19.00){\makebox(0,0)[cc]{$p_b$}}
\put(46.00,30.00){\makebox(0,0)[cc]{$p_b$}}
\put(103.00,79.00){\makebox(0,0)[cc]{$p_c$}}
\put(108.00,35.00){\makebox(0,0)[cc]{$p_c$}}
\put(108.00,35.00){\makebox(0,0)[cc]{$p_c$}}
\put(44.00,14.00){\makebox(0,0)[cc]{$p_c$}}
\end{picture}

\begin{center}
Fig.~1
\end{center}

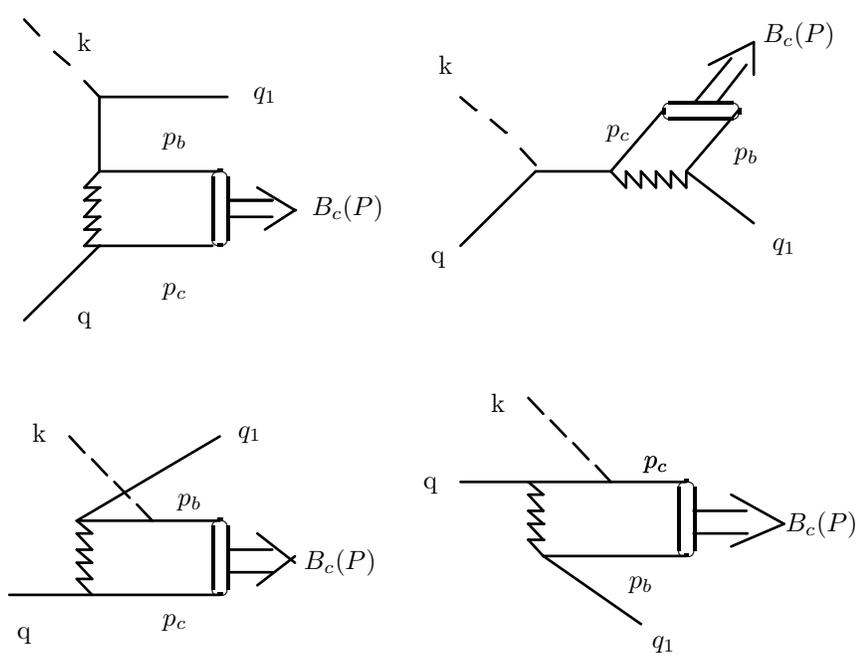
\end{document}